\documentclass{article}
\usepackage[latin9]{inputenc}
\usepackage{amsmath}
\usepackage{amssymb}
\usepackage{graphicx}

\makeatletter
\usepackage{spconf}

\title{Speech Dereverberation in the STFT Domain}
\name{Author(s) Name(s)\thanks{Thanks to XYZ agency for funding.}}
\address{Author Affiliation(s)}

\name{Richard Stanton and Mike Brookes}
\address{\normalsize{Imperial College London}\\
\texttt{\normalsize{\{rs408, mike.brookes\}@imperial.ac.uk}}}
\usepackage[printonlyused,withpage]{acronym}

\usepackage{ifthen}
\providecommand{\nomenclature}{}
\renewcommand{\nomenclature}[2]{\ifthenelse{\equal{#2}{f}}{\acf{#1}}{\ifthenelse{\equal{#2}{s}}{\acs{#1}}{\ifthenelse{\equal{#2}{l}}{\acl{#1}}{\ifthenelse{\equal{#2}{p}}{\acp{#1}}{\ac{#1}}}}}}

\usepackage{url}

\makeatother

\begin{document}
%\ninept
\maketitle 
\begin{abstract}
Reverberation is damaging to both the quality and the intelligibility
of a speech signal. We propose a novel single-channel method of dereverberation
based on a linear filter in the Short Time Fourier Transform domain.
Each enhanced frame is constructed from a linear sum of nearby frames
based on the channel impulse response. The results show that the method
can resolve any reverberant signal with knowledge of the impulse response
to a non-reverberant signal.
\end{abstract}
%\acronymnolist{/Users/Rich/Dropbox/Postgraduate/Research/SAPBibTeX/sapacronyms.txt}

\section{Introduction}

\label{sec:intro}

Speech is inherently non-stationary, therefore speech processing algorithms
are frequently applied to short frames in which the speech is quasi-stationary.
Furthermore, speech is sparse in the time-frequency domain, allowing
us to distinguish and enhance the speech content well. Therefore the
Short Time Fourier Transform (STFT) domain is the domain of choice
for many speech and audio based algorithms.

Reverberation occurs from multi-path propagation of an acoustic signal,
$s[n]$, through a channel with impulse response $h[n]$ to a microphone.
Reverberation causes speech to sound distant and spectrally distorted
which reduces intelligibility \cite{Naylor2005}. The further the
source from the microphone the greater the effects of reverberation.
Automatic speech recognition is severly hindered by reverberation
\cite{Kingsbury1998,Sehr2006}. Beamformers utilise the time difference
of arrival to each sensor in an array to spatially filter a sound
field. Due to the multi-path propagation, beamformers fail in reverberant
environments. Therefore channel inversion methods are of high importance
in spatial filtering fields.

There already exists several dereverberation algorithms in the STFT
domain. For example spectral subtraction has been used to estimate
the power spectrum of the late reverberation and subtract this from
the current spectrum to leave the direct path, \cite{Lebart2001};
this approach was extended in \cite{Habets2004} to introduce the
frequency dependence of the reverberation time.

Other methods of dereverberation exist which utilise knowledge of
the system impulse response, $h[n]$, however now exist in the STFT
domain. Least squares has previously been used to create an inverse
filter from knowledge of the impulse response, \cite{Widrow1984}.
This was extended into the multichannel domain with the Multiple-input/output
INverse Theorem (MINT), \cite{Miyoshi1988}, which is capable of finding
exact inverse filters, through the use of multiple transmission channels.

We wish to create an algorithm in the STFT domain which utilises
knowledge of the impulse response, $h[n]$, for the uses of dereverberation.
However simply creating an inverse filter in the STFT domain is not
straightforward, as the STFT process is time-variant. We present a
single-channel method of dereverberation based on a linear filter
which combines nearby frames which uses a novel method to account
for the time varying nature of the STFT domain. The frames are linearly
combined using coefficients computed through a least squares based
method on the impulse response.

The remainder of the paper is as follows. In Section \ref{sec:Method}
the method is outlined. Section \ref{sssec:OptimalCoefficients} details
the process to select the optimal coefficients for dereverberation.
The results of the algorithm are detailed in Section \ref{sec:Results}
and conclusions are drawn in Section \ref{sec:Conclusions}.

\section{STFT-Domain Dereverberation}

\label{sec:Method}

The observed reverberant signal, $y[n]$, at the microphone is the
convolution of the source signal, $s[n]$, and the channel impulse
response, $h[n]$:
\begin{eqnarray}
y\left[n\right] & = & \sum_{m=0}^{M-1}h[m]s[n-m].\label{eq:convolution}
\end{eqnarray}
Exploiting knowledge of the channel impulse response, we propose a
new method to reduce the effects of reverberation on $y[n]$, to form
an estimate, $\hat{s}[n]$, of the original signal.

The reverberant signal is transformed into the STFT domain using a
window, $w[n]$ and an overlapping factor $Q$:
\begin{align}
Y_{k}\left[l\right]= & \sum_{n=0}^{QR-1}y[n+lR]w[n]e^{-j2\pi\frac{kn}{QR}},\label{eq:STFT}
\end{align}
where $l$ represents the frame number, $k$ the frequency bin and
$R$ the frame increment. The enhanced signal is formed through a
linear sum of nearby frames of the reverberant signal:
\begin{eqnarray}
\widehat{S}_{k}[l] & = & \sum_{r=-A}^{B}G_{k}[r]Y_{k}[l-r],\label{eq:SpecSubtract}
\end{eqnarray}
where $A$ is the number of future frames and $B$ is the number
of past frames to be used in the enhancement. The resulting frames
are then transferred back into time frames with the inverse Discrete
Fourier Transform (DFT):
\begin{align}
\hat{s}[l,m]= & \frac{1}{QR}\sum_{k=0}^{QR-1}\hat{S}_{k}[l]e^{j2\pi\frac{km}{QR}},\label{eq:InverseDFT}
\end{align}
which are then overlap-added \cite{Allen1977b} to form the enhanced
time signal:
\begin{align}
\hat{s}[n]= & \sum_{l=l_{n}-Q+1}^{l_{n}}\hat{s}[l,n-lR]w[n-lR].\label{eq:Overlap-Add}
\end{align}
where $l_{n}=\left\lfloor \frac{n}{R}\right\rfloor $. Perfect reconstruction,
$\hat{s}[n]=y[n]$, is obtained with the coefficients $G_{k}[r]=\delta[r]$
provided that the window used for analysis and synthesis satisfies:
\[
\sum_{q=0}^{Q-1}w^{2}[qR+n]=1\qquad\forall n\in\left[0,\,R-1\right].
\]

\section{Optimal Coefficients}

\label{sssec:OptimalCoefficients}Assuming that $h[n]$ is known,
our goal is to determine the filter coefficients $\mathbf{G}_{k}=\begin{bmatrix}G_{k}[-A] & \ldots & G_{k}[B]\end{bmatrix}^{T}$
so that $\hat{s}[n]\approx s[n]$.

Consider the response of \eqref{eq:SpecSubtract} when the input signal
is an impulse at sample $\lambda$:
\[
s^{(\lambda)}[n]=\delta[n-\lambda],\quad\lambda\in[0,R-1].
\]

When processing in the STFT domain, the earliest output frame that
is affected by the impulse occurs at $l_{min}=1-Q-A$, whereas the
latest frame affected is $l_{max}=1+B+\left\lfloor \frac{M+\lambda-2}{R}\right\rfloor $.
Appling the process from \eqref{eq:SpecSubtract} we can find a relationship
between the channel STFT of the impulse response, $H_{\lambda}[l,k]$,
and the desired impulse response $\tilde{H}_{\lambda}[l,k]$, which
is the STFT of the direct path impulse response, when there are no
reflections present.

\begin{figure}
\begin{centering}
\includegraphics[width=1\linewidth]{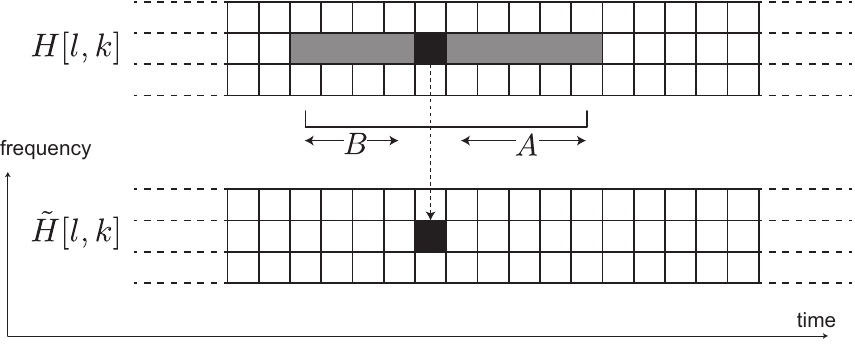}
\par\end{centering}

\caption{\label{fig:ImpulseMapping-2}The above plots show the STFT of both
$H[l,k]$ and $\tilde{H}[l,k]$. For each frequency bin the filter
linearly combines future and past frames of $H[l,k]$ to best match
$\tilde{H}[l,k]$.}
\end{figure}

We determine $\mathbf{G}_{k}$ to minimise the difference between
the two. So for each frequency bin, $k$, we have an overdetermined
set of equations:
\begin{align}
\hat{H}^{(\lambda)}[l,k;\mathbf{G_{k}}]= & \sum_{r=A}^{B}G_{k}[r]H^{(\lambda)}[l-r,k]\approx\tilde{H}^{(\lambda)}[l,k],\label{eq:SpecSubImpulse-3}
\end{align}
for each $\lambda=[0:R-1]$ and $l=\left[l_{min}:l_{max}\right]$.
This gives us $\left(2+A+B+Q\right)R+M-1$ equations, with $A+B+1$
unknowns. This process is shown in Fig. \ref{fig:ImpulseMapping-2}.
We combine $B$ past frames with $A$ future frames to best approximate
the current frame from the desired impulse response.

We solve these equations using linear least squares, \cite{Lawson1987},
to find:
\begin{align}
\mathbf{G}_{k}=\arg\min_{\mathbf{G}_{k}} & \sum_{\lambda=0}^{R-1}\sum_{l=l_{min}}^{l_{max}}\left(\hat{H}^{(\lambda)}[l,k;\mathbf{G_{k}}]-\tilde{H}^{(\lambda)}[l,k]\right)^{2}.\label{eq:CoefficientsMinimisationProb-2}
\end{align}

The overall impulse response of the computed channel is time-variant
but we can determine an average channel response as the inverse STFT
of:
\begin{align}
\hat{H}\left[l,k\right]= & \frac{1}{R}\sum_{\lambda=0}^{R-1}\hat{H}^{(\lambda)}[l,k;\mathbf{G_{k}}]\exp\left(2\pi k\frac{\lambda}{QR}\right),\label{eq:EffectChannelResponse-2}
\end{align}
where a phase shift is applied to correspond with the sample position
within the frame.

\subsection{Time domain error bound}

The above minimisation problem minimises the reverberation present
in the enhanced signal. Let us define the error in the impulse responses
in both the time domain and the STFT domain as:
\[
h_{e}[n]=\tilde{h}[n]-\hat{h}[n].
\]
The error in a single frame in the STFT domain is as follows:
\begin{eqnarray*}
H_{e,k}[l] & = & \tilde{H}_{k}^{(\lambda)}\left[l\right]-\hat{H}_{k}^{(\lambda)}\left[l\right].
\end{eqnarray*}
The total power of the error in the STFT domain across all frames,
frequencies and shifts is denoted: 
\[
P_{f}[n]=\frac{1}{QR}\sum_{k=0}^{QR-1}\sum_{\lambda=0}^{R-1}\sum_{l=l_{min}}^{l_{max}}\left|H_{e}[l,k]\right|^{2}.
\]
Using Parseval's theorem, the power of the error in the time domain
is given as:
\begin{equation}
\sum_{n=0}^{QR-1}\left|h_{e}[l,n]\right|^{2}=\frac{1}{QR}\sum_{k=0}^{QR-1}\sum_{\lambda=0}^{R-1}\sum_{l=l_{min}}^{l_{max}}\left|H_{e}[l,k]\right|^{2}.\label{eq:STFTerror-1}
\end{equation}
Alternatively we can express the error power, in the time domain,
as the weighted sum of the frames with the window function:
\[
h_{e}[lR+n]=\sum_{q=0}^{Q-1}w[qR+n]h_{e}[qR+n,l-q].
\]
We sum over all time samples to give the total error:
\begin{gather}
\sum_{l=0}^{N-1}\sum_{n=0}^{R-1}h_{e}^{2}[lR+n]=\qquad\qquad\label{eq:TimeError-1}\\
\sum_{l}\sum_{n=0}^{R-1}\left(\sum_{q=0}^{Q-1}w[qR+n]h_{e}[qR+n,l-q]\right)^{2}.\nonumber 
\end{gather}
Thus applying the Cauchy Schwatz inequality to \eqref{eq:STFTerror-1}
and \eqref{eq:TimeError-1}, we can show that the error in the STFT
domain is an upper bound for the time domain error:
\begin{gather*}
\sum_{l=0}^{N-1}\sum_{n=0}^{R-1}\left(\sum_{q=0}^{Q-1}w[qR+n]h_{e}[qR+n,l-q]\right)^{2}\\
\leq\frac{1}{QR}\sum_{k=0}^{QR-1}\sum_{\lambda=0}^{R-1}\sum_{l=0}^{N-1}\left|H_{e}[l,k]\right|^{2}.
\end{gather*}
Therefore solving the related problem in the STFT domain places an
upper bound on the amount of reverberation in our output signal.

\section{Evaluation}

\label{sec:Results}To evaluate the reduction in reverberation, we
use two metrics: the Direct-to-Reverberant Ratio (DRR) \cite{Bronkhorst1999}
and the Signal-to-Reverberation Ratio (SRR) \cite{Quackenbush1988}.
To evaluate the perceptual quality of the enhanced signals Perceptual
Evaluation Of Speech Quality (PESQ), \cite{ITU_T_P862_a}, is used.
The DRR $\left[\mathrm{dB}\right]$ is defined as follows:
\begin{eqnarray}
\mathrm{DRR} & = & \frac{10}{R}\sum_{\lambda=0}^{R-1}\log_{10}\left\{ \frac{E_{d}\left(\lambda\right)}{\left(\sum_{n}h_{\lambda}^{2}[n]\right)-E_{d}\left(\lambda\right)}\right\} ,\label{eq:DRR-4}
\end{eqnarray}
where $E_{d}$ is the direct path energy. The direct path in the impulse
response may occur in between samples, therefore the path energy will
be spread across the nearby samples with a sinc function. Thus the
direct path energy is computed using a convolution with a sinc function
with a varying offset until a maximum is found:
\[
E_{d}\left(\lambda\right)=\max_{\sigma}\sum_{n=-\eta}^{\eta}\left(\frac{\sin\left(\pi\left(n+\sigma\right)\right)}{\pi\left(n+\sigma\right)}h_{\lambda}[n+n_{d}]\right)^{2},
\]
where $n_{d}$ is the nearest index of the direct path in the impulse
response, $\eta=8$ is the number of sidelobes of the sinc function
to use in the summation and $\sigma=[-1:1]$ is the offset that finds
the maximum power.

The SRR $\left[\mathrm{dB}\right]$ is defined on a frame by frame
basis and then averaged across the whole signal:
\begin{equation}
\mathrm{SRR}_{\mathrm{seg}}=\frac{10}{M}\sum_{k=0}^{M-1}\log_{10}\left\{ \frac{\sum_{n=kR}^{kR+QR-1}s_{d}[n]^{2}}{\sum_{n=kR}^{kR+QR-1}\left(s_{d}[n]-\hat{s}[n]\right)^{2}}\right\} ,\label{eq:SRR-2}
\end{equation}
where $M$ is the total number of frames, $s_{d}[n]$ represents the
orignal direct path signal and $\hat{s}[n]$ is the enhanced signal.
It gives a measure of the reverberation power in relation to the useful
direct path. It is a similar measure to the DRR but uses speech signals
rather than the channel response.

The optimal coefficients from Section \ref{sssec:OptimalCoefficients}
were calculated for a Room Impulse Response (RIR) and the corresponding
channel response from \eqref{eq:EffectChannelResponse-2} was found.
A total of 600 RIRs were used to test the system. These correspond
to a single source and microphone in 40 different rooms and 15 different
position combinations in each. The impulse responses were generated
using the Room Impulse Response Generator\emph{ }from \cite{Habets2006},
which is based on the image method \cite{Allen1979}. In all cases
$Q=4$, $R=64$, $A=9$, $B=9$.

As both the SRR and PESQ work on speech samples the TIMIT core test
set \cite{Garofolo1993} was chosen. Each speech sample was convolved
with each $h[n]$ before undergoing enhancement as described in \eqref{eq:SpecSubtract}.
The before and after signals, $y[n]$ and $\hat{s}[n]$, were then
used with the SRR and PESQ metrics to gauge any improvement.

The performance of the proposed algorithm has been compared to the
time domain inverse filter as proposed by Widrow, \cite{Widrow1984}.
The method designs an inverse filter, $g[n]$, through least squares
to best invert the system response, $h[n]$, \cite{Miyoshi1988}:
\begin{align*}
\begin{bmatrix}\vdots\\
0\\
1\\
0\\
\vdots
\end{bmatrix}= & \begin{bmatrix}h\left[0\right] &  &  & 0\\
\vdots & h\left[0\right]\\
h\left[N_{h}-1\right] & \vdots & \ddots\\
 & h\left[N_{h}-1\right] &  & h\left[0\right]\\
 &  & \ddots & \vdots\\
0 &  &  & h\left[N_{h}-1\right]
\end{bmatrix}\\
 & \quad\times\begin{bmatrix}g\left[0\right]\quad g\left[1\right]\quad\ldots\quad g\left[M-1\right]\end{bmatrix}^{T},
\end{align*}
where $N_{h}=1024$  in our case.

\subsection{Results}

\begin{figure}[t]
\begin{centering}
\includegraphics[width=1\linewidth]{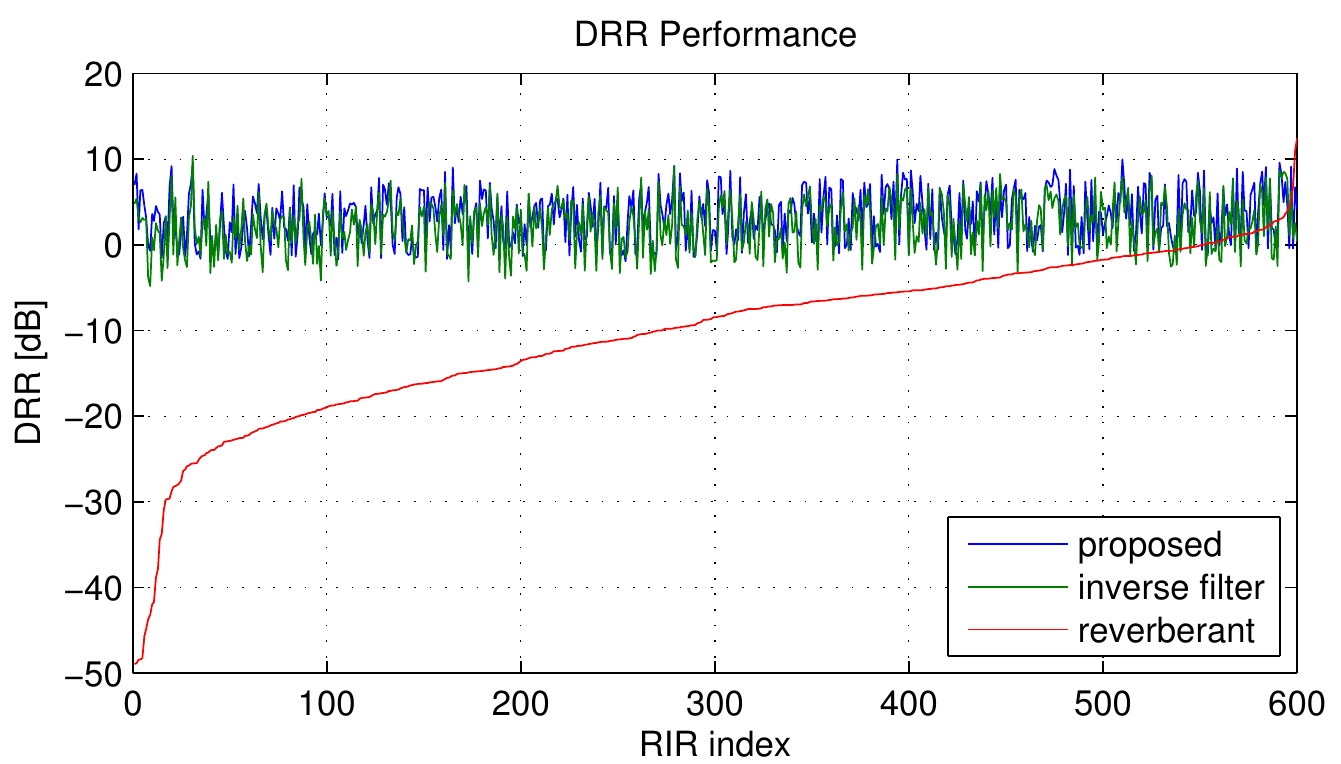}
\par\end{centering}

\begin{centering}

\par\end{centering}

\caption{\label{fig:DRR Improvement}The DRR after the algorithm for 600 RIRs
shows a sizeable improvement over the initial reverberant signal,
mean average improvement of $1.0\,\mathrm{dB}$.}
\end{figure}

The DRR was computed for both $h[n]$ and $\hat{h}[n]$ across all
600 RIRs. The results comparing the DRR before and after the algorithm
are shown in Fig.~\ref{fig:DRR Improvement}.
The DRR improved for all the impulse responses tested except those
where the original DRR exceed $0$~dB. The resulting performance
is independent of the amount of reverberation in the initial signal
and hovers close to $6\,\mathrm{dB}$, giving an improvement of up
to $34\,\mathrm{dB}$. Thus the algorithm is able to reduce reverberation
to the same level regardless of how reverberant the original channel
is.

\begin{figure}[t]
\begin{centering}
\includegraphics[width=1\linewidth]{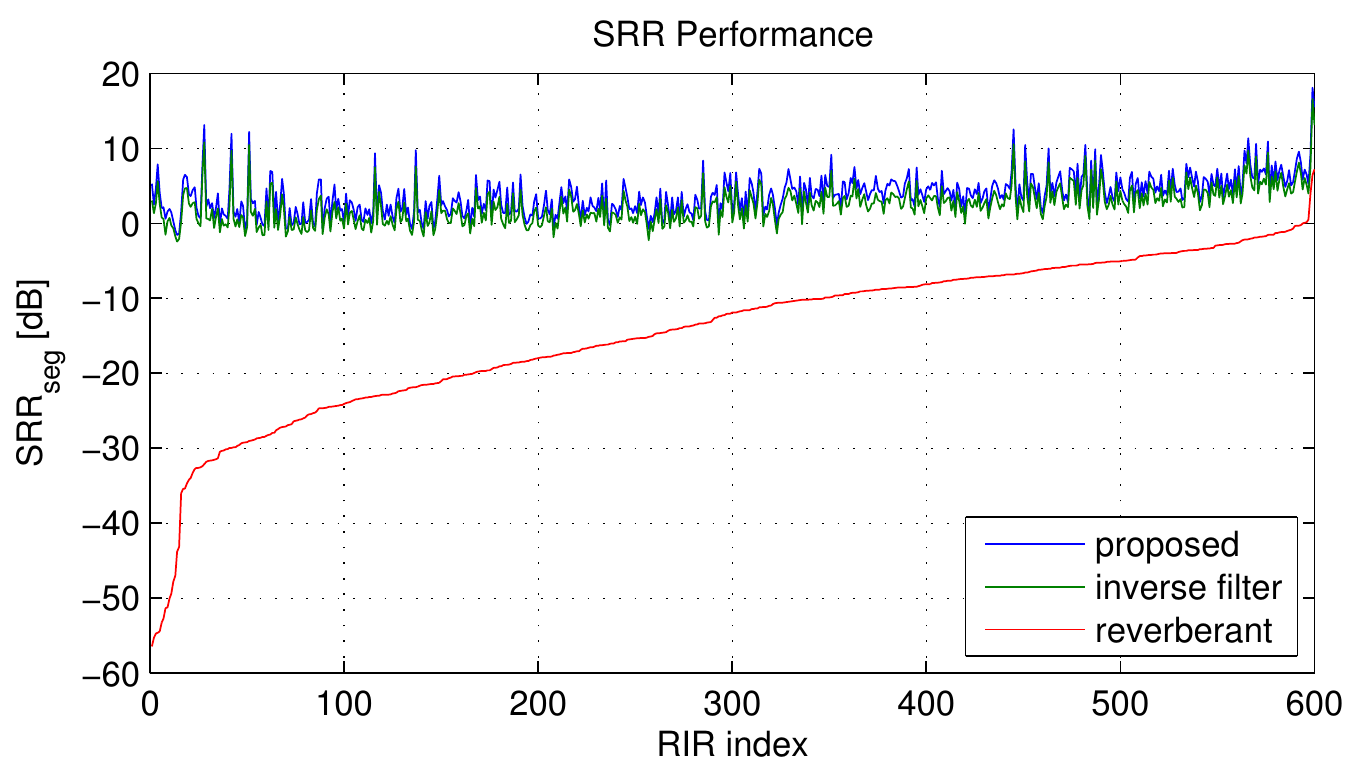}
\par\end{centering}

\begin{centering}
\includegraphics[width=1\linewidth]{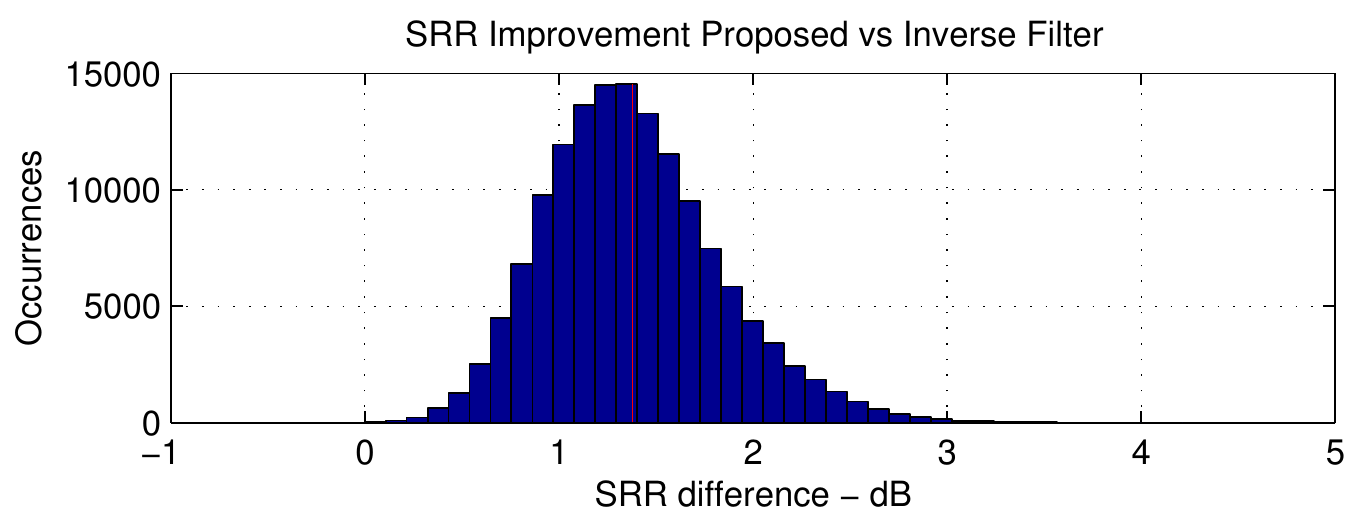}
\par\end{centering}

\caption{\label{fig:SRR Improvement}The speech signals after enhancement show
a much improved SRR compared to the reverberant signals, mean average
improvement of $1.4\,\mathrm{dB}$.}
\end{figure}

The averaged SRR for each RIR is shown in Fig.~\ref{fig:SRR Improvement}.
It follows a similar pattern to the DRR. The enhanced signals hover
around $0\,\mathrm{dB}$.
 When the original SRR surpassed $0\,\mathrm{dB}$, the algorithm
was unable to make any further improvements, and caused slight degredation
to these non-reverberant signals.

\begin{figure}
\begin{centering}
\includegraphics[width=1\linewidth]{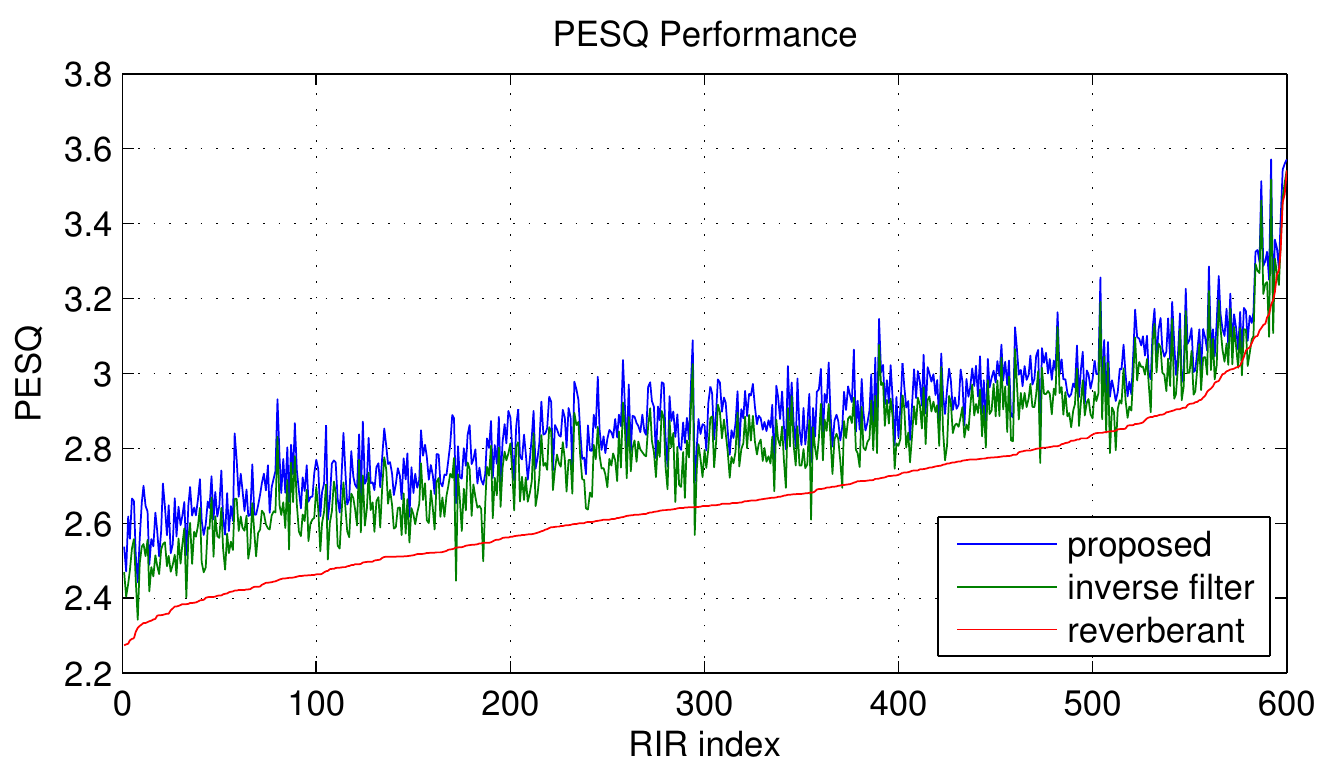}
\par\end{centering}

\caption{\label{fig:PESQ Improvement}PESQ is shown for 600 different RIRs
before and after enhancement. Each point is the average of 240 utterances
for that RIR, mean average improvement of $0.08\,\mathrm{PESQ}$.}
\end{figure}

The averaged PESQ results are shown in Fig.~\ref{fig:PESQ Improvement}.
 The enhancement gave a small gain in perceptual quality which, whilst
it does not show the removal of reverberation, does show that the
algorithm does not introduce significant distortion. Due to the limited
improvement in the perceived speech quality the algorithm has good
uses in approaches which require signals without reverberation, rather
than end user perceptual improvements.

Samples of the reverberant and processed speech are available on the
internet: \cite{Stanton2013}.

\section{Conclusions}

\label{sec:Conclusions}

We have described a novel approach to dereverberation using a linear
filter in the STFT domain. Using knowledge of the channel impulse
response we can find an optimal combination of frames to reduce the
effects of reverberation. The algorithm gives clear performance gains
in dereverberation. Both the DRR and the SRR show that regardless
of the amount of initial reverberation present, the enhanced signal
has a similar low level of reverberation present, whilst not introducing
distortion.

We have shown that the proposed STFT domain algorithm is as good as
the time domain inverse filter; allowing us to apply dereverberation
in the more appropriate domain without loss of performance.

We have overcome the time-variance of the STFT by considering all
the possible impulse positions within a single frame.

By working in the STFT domain we can solve each frequency band, $k$,
independently. The above give a useful framework that suits many applications
already processing in this domain.\clearpage{}\bibliographystyle{IEEEbib}
\bibliography{/Users/Rich/Dropbox/Postgraduate/Research/SAPBibTeX/sapref,/Users/Rich/Dropbox/Postgraduate/Research/SAPBibTeX/MyReferences}

\begin{thebibliography}{10}

\bibitem{Naylor2005}
P.~A. Naylor and N.~D. Gaubitch,
\newblock ``Speech dereverberation,''
\newblock in {\em Proc. Intl. Workshop Acoust. Echo and Noise Control
  ({IWAENC})}, Eindhoven, The Netherlands, Sept. 2005.

\bibitem{Kingsbury1998}
Brian E.~D. Kingsbury, Nelson Morgan, and Steven Greenberg,
\newblock ``Robust speech recognition using the modulation spectrogram,''
\newblock {\em Speech communication}, vol. 25, no. 1, pp. 117--132, 1998.

\bibitem{Sehr2006}
A.~Sehr, M.~Zeller, and W.~Kellermann,
\newblock ``Hands-free speech recognition using a reverberation model in the
  feature domain,''
\newblock in {\em Proc. European Signal Processing Conf. (EUSIPCO)}, Florence,
  Italy, Sept. 2006.

\bibitem{Lebart2001}
K.~Lebart, J.~M. Boucher, and P.~N. Denbigh,
\newblock ``A new method based on spectral subtraction for speech
  de-reverberation,''
\newblock {\em Acta Acoustica}, vol. 87, pp. 359--366, 2001.

\bibitem{Habets2004}
E.~A.~P. Habets,
\newblock ``Single-channel speech dereverberation based on spectral
  subtraction,''
\newblock in {\em Proc. Workshop Circuits, Systems and Signal Processing
  ({ProRISC})}, Veldhoven, The Netherlands, Nov. 2004, pp. 250--254.

\bibitem{Widrow1984}
Bernard Widrow and Eugene Walach,
\newblock ``Adaptive signal processing for adaptive control,''
\newblock in {\em Proc. {IEEE} Intl. Conf. on Acoustics, Speech and Signal
  Processing ({ICASSP})}. IEEE, 1984, vol.~9, pp. 191--194.

\bibitem{Miyoshi1988}
M.~Miyoshi and Y.~Kaneda,
\newblock ``Inverse filtering of room acoustics,''
\newblock {\em {IEEE} Trans. Acoust., Speech, Signal Process.}, vol. 36, no. 2,
  pp. 145--152, Feb. 1988.

\bibitem{Allen1977b}
J.~Allen and L.~Radiner,
\newblock ``A unified approach to short-time {Fourier} analysis and
  synthesis,''
\newblock {\em Proc. {IEEE}}, vol. 65, no. 11, pp. 1558--1564, 1977.

\bibitem{Lawson1987}
Charles~L Lawson and Richard~J Hanson,
\newblock {\em Solving least squares problems}, vol. 161,
\newblock SIAM, 1987.

\bibitem{Bronkhorst1999}
Adelbert~W Bronkhorst and Tammo Houtgast,
\newblock ``Auditory distance perception in rooms,''
\newblock {\em Nature}, vol. 397, no. 6719, pp. 517--520, 1999.

\bibitem{Quackenbush1988}
Schuyler~R. Quackenbush, Thomas~P. Barnwell, III, and Mark~A. Clements,
\newblock {\em Objective Measures of Speech Quality},
\newblock Prentice Hall, Jan. 1988.

\bibitem{ITU_T_P862_a}
ITU-T,
\newblock ``Perceptual evaluation of speech quality ({PESQ}), an objective
  method for end-to-end speech quality assessment of narrowband telephone
  networks and speech codecs,''
\newblock Recommendation P.862, International Telecommunications Union
  ({ITU-T}), Feb. 2001.

\bibitem{Habets2006}
E.~A.~P. Habets,
\newblock ``Room impulse response generator,''
\newblock Tech. {R}ep., Technische Universiteit Eindhoven, 2006.

\bibitem{Allen1979}
J.~B. Allen and D.~A. Berkley,
\newblock ``Image method for efficiently simulating small-room acoustics,''
\newblock {\em J. Acoust. Soc. Am.}, vol. 65, no. 4, pp. 943--950, Apr. 1979.

\bibitem{Garofolo1993}
John~S. Garofolo, Lori~F. Lamel, William~M. Fisher, Jonathan~G. Fiscus,
  David~S. Pallett, Nancy~L. Dahlgren, and Victor Zue,
\newblock ``{TIMIT} acoustic-phonetic continuous speech corpus,''
\newblock Corpus LDC93S1, Linguistic Data Consortium, Philadelphia, 1993.

\bibitem{Stanton2013}
Richard Stanton,
\newblock ``Speech dereverberation in the {STFT} domain,''
  \url{http://www.commsp.ee.ic.ac.uk/~sap/people/richard-stanton/}, Nov. 2013.

\end{thebibliography}

\end{document}